\documentclass[preprint,showpacs,preprintnumbers,amsmath,amssymb]{revtex4}

\usepackage{graphicx}
\usepackage{latexsym}
\usepackage{amsmath}
\usepackage{color}

\newcommand{\bqu}{\begin{equation}}
\newcommand{\udarrow}[2]{\smash{\mathop{%
  \hbox to 0.4cm{$\rightleftharpoons$}}\limits^{#1}\limits_{#2}}}

\begin{document}

\preprint{}

\title{
Rate Equation Approaches to 
Amplification of Enantiomeric Excess
and Chiral Symmetry Breaking
}

\author{Yukio Saito}
\author{Hiroyuki Hyuga}

\affiliation{
Department of Physics, Keio University, Yokohama 223-8522, Japan
}


\begin{abstract}
Theoretical models and rate equations relevant to the Soai reaction are reviewed. It is found that in a production of chiral molecules from an achiral substrate autocatalytic processes can induce either enantiomeric excess (ee) amplification or chiral symmetry breaking. Former terminology means that the final ee value is larger than the initial value but depends on this, whereas the latter means the selection of a unique value of the final ee, independent of the initial value. The ee amplification takes place in an irreversible reaction such that all the substrate molecules are converted to chiral products and the reaction comes to a halt. The chiral symmetry breaking is possible when recycling processes are incorporated. Reactions become reversible and the system relaxes slowly to a unique final state. The difference between the two behaviors is apparent in the flow diagram in the phase space of chiral molecule concentrations. The ee amplification takes place when the flow terminates on a line of fixed points (or a fixed line), whereas symmetry breaking corresponds to the dissolution of the fixed line accompanied by the appearance of fixed points. Relevance of the Soai reaction to the homochirality in life is also discussed. 
\end{abstract}

\maketitle

\section{Introduction}
In the Human Genome Project (from the year 1990 to 2003) 
sequences of chemical base pairs that make up human DNA
were intensively analyzed and determined, 
as it carries important genetic information.
DNA is a polymer made up of a large number of deoxyribonucleotides, 
each of which 
composed of a nitrogenous base, 
a sugar and one or more phosphate groups
 \cite{stryer98}. 
Similar ribonucleotides polymerize to form RNA,
which is also an important substance to produce a 
template for protein synthesis.
RNA sometimes carries genetic information but rarely shows enzymatic functions 
 \cite{stryer98}.

One big issue of the post Genome Project is the proteomics since
proteins play crucial roles in virtually all
biological processes such  as
enzymatic catalysis, coordinated motion,
 mechanical support, etc. \cite{stryer98}.
A protein is a long polymer of amino acids, and folds into
regular structures to show its biological function.

Sugars and amino acids in these biological polymers
contain carbon atoms, each of which is 
connected to four different groups, and consequently is able to take
two kinds of stereostructures.
The two stereoisomers are mirror images or enantiomers
 of each other, and  are called the {\small D}- and {\small L}-isomers. 
Two isomers should have the same physical properties except 
optical responses to polarized light. Therefore, a simple symmetry argument
leads to the conclusion that there are equal amount of {\small D}- and 
{\small L}-amino acids or sugars in life.
But the standard  biochemical textbook \cite{stryer98} tells us that
"only {\small L}-amino acids are constituents of proteins"
( \cite{stryer98}, p.17), and
"nearly all naturally occurring sugars belong to the {\small D}-series"
( \cite{stryer98}, p.346).
There is no explanation why the chiral symmetry is broken in life
and how the homochirality has been brought about on earth.

It is Pasteur who first recognized the chiral symmetry breaking
in life in the middle of the 19th century. 
By crystallizing optically inactive sodium anmonium racemates,
he separated two enantiomers of sodium ammonium 
tartrates with opposite optical activities by
means of their asymmetric crystalline shapes  \cite{pasteur849}.
Since the activity is observed even
in the solution, 
it is concluded that the optical activity is
due to the molecular asymmetry or chirality,
not due to the crystalline symmetry.
Because two enantiomers with different chiralities
are identical in every chemical and physical
properties except the optical activity, 
"artificial products have no molecular asymmetry"
and
in 1860 Pasteur stated that 
"the molecular asymmetry of natural organic products"
establishes "the only well-marked line of demarcation that can
at present be drawn between the chemistry of dead matter and the 
chemistry of living matter"  \cite{japp898}.
Also by using the fact that asymmetric chemical agents react differently 
with two types of enantiomers, he separated enantiomers by
fermentation.
Therefore, once one has an asymmetric substance,
further separation of two enantiomers or even a production of a single
type of enantiomer may follow \cite{noyori02}.
But how has the first asymmetric organic compound 
been chosen 
in a prebiotic world?

This problem of the origin of homochirality in life has attracted 
attensions of many scientists in relation to the origin of life itself
since the discovery of Pasteur
\cite{japp898,calvin69,goldanskii+88,gridnev06}.
Japp expected a "directive force" when "life first arose"
\cite{japp898}, 
and various "directive forces" are proposed later such as 
different intensities of circularly polarized
light in a primordial era, adsorption on optically active crystals,
or the parity breaking in the weak interaction.
However, the expected degree of chiral asymmetry or the 
value of the enantiomeric excess (ee) turns out to be very small
\cite{goldanskii+88}, and 
one needs a mechanism to amplify ee enormously to a level of homochirality.

Another scenario for the origin of homochirality
was suggested by Pearson  \cite{peason898} 
such that the chance breaks the chiral symmetry.
Though the mean number of right- and left-handed enantiomers are the
same, there is nonzero probability of deviation from the
equal populations of both enantiomers. 
The probability to establish homochirality 
in a macroscopic system is, of course, very small \cite{mills32},
but "chance produces a slight majority of one type of" enantiomers
and "asymmetric compounds when they have once arisen" act
as "breeders, with a power of selecting of their own kind of asymmetry
form" \cite{peason898}.
In this scenario, produced enantiomer acts as a chiral catalyst
for the production of its own kind 
 and hence this process should be autocatalytic. 

Nearly a century later, 
 a long sought autocatalytic system 
with spontaneous amplification of chiral asymmetry
is found by Soai and his coworkers \cite{soai+95}:
In a closed reactor with achiral substrates 
addition of a small amount of chiral products with a slight
enantiomeric imbalance
yields the final products with
an overwhelmingly amplified ee
\cite{soai+95}.
The autocatalytic system also shows significant ee amplification
under a variety of organic and inorganic chiral initiators with 
a small enantiomeric imbalance  \cite{soai+00}.
When many reaction runs are performed without chiral additives,
about half of the runs end up with the majority of one enantiomer,
and the other half end up with the opposite enantiomer.
The probability distribution of the ee is bimodal with
double peaks, showing amplification of ee and further indicating the
occurrence of chiral symmetry breaking
\cite{soai+03,gridnev+03,singleton+03}.

As for the theoretical research on the chiral symmetry breaking,
Frank is the  first to show that a linear autocatalysis
with an antagonistic nonlinear chemical reaction 
can lead to homochirality \cite{frank53}.
His formulation with rate equations
corresponds to the mean-field analysis of the
phase transition in nonequilibrium situation
\cite{landau+54}, and 
other variants have been proposed 
\cite{goldanskii+88,avetisov+96,girard+98,kondepudi+01,todd02,
iwamoto02,iwamoto03}.
All these analyses are carried out 
 for
 open systems where the concentration of an achiral substrate is kept constant.
Asymptotically the system 
 approaches a unique steady state 
\cite{goldanskii+88,avetisov+96,girard+98,kondepudi+01,todd02}
 or even an oscilatory state
 \cite{iwamoto02,iwamoto03},
independent of the initial condition.
When the final state is chirally asymmetric,
one may call this a chiral symmetry breaking in the sense of
statistical mechanics,
and it is indicated in the right upper corner in Table 1.
On the other hand, the Soai reaction is performed
in a closed system and 
 an achiral substrate is converted to
chiral products irreversibly:
 The substrate concentration decreases in time, 
and eventually the reaction comes to a halt.
After the reaction the ee value  increases
 but its final value is found to depend on the initial state.
Even though the ee is amplified, the history dependent behavior
is different 
from what we expect from phase transitions in statistical mechanics,
so that we 
write down  "ee amplification"  in
the left lower corner in Table 1.
We have analyzed theoretically the chiral symmetry breaking in a closed system 
in general, and found that with an irreversible 
 nonlinear autocatalysis the system may show ee amplification 
 as in the Soai reaction.
Furthermore, with an additional recycling back reaction,
a unique final state with a finite ee value is found possible,
which we call a chiral symmetry breaking as is shown in 
the right lower corner in Table 1
\cite{saito+04a,saito+04b,saito+05a,saito+05b,saito+05c,shibata+06}.
There are also many theoretical works on the Soai reaction in a closed
system 
\cite{sato+01,sato+03,blackmond+01,buhse03,islas+05,lente04,lente05}.
Here we give a brief survey of various theoretical
models of the ee amplification
and the chiral symmetry breaking in a closed system, 
though it
is by no means exhaustive.

\newcommand{\lw}[1]{\smash{\lower1.7ex\hbox{#1}}}
\begin{table}[htb]
\caption{
The ee amplification and the chiral symmetry breaking in an open and a closed system.
}
\label{tab1}
\begin{center}
\renewcommand{\arraystretch}{1.7}
\begin{tabular}{|c|c@{\quad\vrule width 1pt}c|c|}
\cline{3-4}
\multicolumn{2}{l@{\vrule width 1pt}}{}&
\multicolumn{2}{c|}{Initial Condition}\\
\cline{3-4}
\multicolumn{2}{l@{\vrule width 1pt}}{}
&Dependent&Independent\\ 
\noalign{\hrule height 1pt}
\lw{System} & Open &&Chiral Symmetry Breaking\\ 
\cline{2-4} &Closed&\quad ee Amplification&Chiral Symmetry Breaking\\
 \hline
\end{tabular}
\end{center}
\end{table}

\section{
Amplification in Monomeric Systems
}

Two stereostructural isomers are called {\footnotesize D}- 
and {\footnotesize L}-enantiomers 
for sugars and amino acids, 
but, for general organic compounds, $R$ and $S$ representation
is in common. 
We adopt $R-S$ representation hereafter.

We consider a production of chiral enantiomers $R$ and $S$ from an achiral 
substrate $A$ 
in a closed system. 
Actually, in the Soai reaction, chiral molecules are produced
by the reaction of two achiral reactants $A$ and $B$
as $A+B \rightarrow R$ or $A+B\rightarrow S$. But in a closed system
a substrate of smaller amount  controls the reaction, 
 since it is first consumed up and
holds the reaction to proceed further. Therefore, in order
to grasp main features of chirality selection, it is sufficient to
assume that an achiral substrate $A$ of a smaller amount
converts to $R$ or $S$. 
The process may further involve formation of
intermediate complexes, oligomers etc., but 
for the purpose to discuss
about relations and difference between the amplification and the
symmetry breaking 
in chirality, we restrict our consideration in this section to
 the simplest case where only monomers are involved.

\subsection{Reaction Schemes}

The spontaneous production of chiral molecules $R$ or $S$ 
from an achiral substrate $A$ is described by reactions 
\begin{align}
A \stackrel{k_0}{\rightarrow} R, \qquad
A \stackrel{k_0}{\rightarrow} S .
\label{eq01}
\end{align} 
We assume the same reaction rate $k_0$ for the nonautocatalytic
spontaneous production of $R$ and $S$  enantiomers, since both of them 
are identical in every chemical and physical aspects:
No advantage factor \cite{goldanskii+88} is assumed.

Since the spontaneous production (\ref{eq01}) yields only a
racemic mixture of two enantiomers, one
has to assume some autocatalytic processes.
The simplest is a linear autocatalysis 
 with a reaction coefficient $k_1$ as 
\begin{align}
A+R \stackrel{k_1}{\rightarrow} 2R, \qquad
A+S \stackrel{k_1}{\rightarrow} 2S .
\label{eq02}
\end{align} 
There may be a linear cross catalytic reaction or erroneous linear catalysis
with a coefficient $k_1'$  as
\begin{align}
A+R \stackrel{k_1'}{\rightarrow} S+R, \qquad
A+S \stackrel{k_1'}{\rightarrow} R+S .
\label{eq03}
\end{align} 
As will be discussed later, the linear autocatalysis alone
is insufficient 
 not only to break the chiral symmetry but also to amplify ee.

Beyond these linear autocatalyses, nonlinear effects 
such as quadratic autocatalysis have been considered as
\cite{goldanskii+88,sato+03,saito+04a};
\begin{align}
A+2R \stackrel{k_2}{\rightarrow} 3R, \qquad
A+2S \stackrel{k_2}{\rightarrow} 3S , 
\label{eq04}
\intertext{and 
cross catalysis as}
A+R+S \stackrel{k_2'}{\rightarrow} 2R+S , \qquad
A+R+S \stackrel{k_2'}{\rightarrow} R+2S .
\label{eq05}
\end{align} 
These higher order autocatalytic processes may actually
be brought about by
the dimer catalysts \cite{girard+98}, but
we confine ourselves to the simplest description 
(\ref{eq04}) and 
(\ref{eq05}) in terms of only monomers $R$ and $S$ in this section.
Consideration on dimers is postponed in 
the following sections.
The reaction (\ref{eq04})
alone can give rise to the amplification
of chirality as will be discussed.
One notes that all these processes (\ref{eq01}) to (\ref{eq05}) 
unidirectionally produce chiral enantiomers
and thus the reaction comes to a halt when the whole 
achiral substrate is consumed; the total process is
irreversible.

\subsection{Rate equations and an order parameter}

Concentrations $r$ and $s$ of two enantiomers $R$ and $S$ vary
according to the reaction processes described in the previous subsection.
In an open system, the achiral substrate is steadily 
supplied in such a way that its concentration $a$ is kept constant.
On the contrary, in a closed system,  
achiral substrates are transformed to chiral products and only
the total concentration $c=a+r+s$ of reactive chemical species 
is kept constant.
By denoting the total concentration of monomeric chiral products 
with a subscript 1 as
\begin{align}
q_1=r+s ,
\label{eq06}
\end{align}
the concentration of the substrate $a$ is determined as
\begin{align}
a=c-q_1  ,
\label{eq07}
\end{align}
which should be non-negative.
Then,  the development of reactions in a closed system can be depicted 
as a flow in the triangular region in the $r$-$s$ phase space 
\begin{align}
0 \le r,s, r+s \le c .
\label{eq08}
\end{align}

The reaction schemes (\ref{eq01}) to (\ref{eq05}) induce
rate equations for $r$ and $s$ as;
\begin{align}
\frac{d r}{dt} = [f(r)+k_1's+k_2'rs]a
\nonumber  \\
\frac{d s}{dt} = [f(s)+k_1'r+k_2'rs]a
\label{eq09}
\end{align}
with an effective rate coefficient
\begin{align}
f(r)=k_0+k_1r+k_2r^2 .
\label{eq10}
\end{align}
Since the right-hand side of Eq.(\ref{eq09}) for velocities 
$\dot r=dr/dt$ and $\dot s=ds/dt$ are always non-negative
and in proportion to the concentration of 
the achiral substrate $a$, 
both $r$ and $s$ never decrease and stop to grow after $a$ 
vanishes ($a=0$). 

The asymptotic behavior of a two-dimensional autonomous dynamical system
is, in general, known to have fixed points or lines where $\dot r=\dot s=0$
or to have limit cycles where $\dot r,~\dot s \ne 0$.
Since the system is irreversible as the concentrations of chiral products 
always increase at the cost of the substrate $A$ as
\begin{align}
\frac{da}{dt}=-[f(r)+f(s)+k'_1(r+s)+2k_2'rs]a \le 0 
\label{eq11}
\end{align}
and no epimerization process $R \leftrightarrow S$ is assumed,
limit cycle is impossible: Production stops when all the
substrate molecules are transformed to chiral ones; $a(t=\infty)=0$.
Since $\dot r=\dot s=0$ for $a=0=c-r-s$, the diagonal line $r+s=c$
is a fixed line for this irreversible system. 
Any points on the fixed line
are marginally stable because they have no guaranteed 
stability along the fixed line.

Our interest is the dynamical behavior of ee
or the chiral symmetry breaking
and, in particular, as to whether ee increases asymptotically.
In order to quantify the 
monomeric ee
or the degree of chiral symmetry breaking, 
we define a monomeric order parameter $\phi_1$ as
\begin{align}
\phi_1= \frac{r-s}{r+s} = \frac{r-s}{q_1} ,
\label{eq12}
\end{align}
which is zero for a symmetric or racemic state ($r=s$), and is nonzero
for a symmetry-broken or chiral state.
The evolution equation of $\phi_1$ is immediately derived 
from Eq.(\ref{eq09}) as
\begin{align}
\frac{d \phi_1}{dt} = A \phi_1- B \phi_1^3
\label{eq13}
\end{align}
with coefficients 
\begin{align}
A(t)&= -2\frac{a}{q_1}(k_0+ k_1' q_1)+B(t), 
\nonumber \\
B(t)&=\frac{1}{2}(k_2-k_2')q_1a.
\label{eq14}
\end{align}
However, as $a(t =\infty)=0$, both the coefficients $A$ and $B$
vanish asymptotically, and one cannot determine the final value
of ee from the evolution equation (\ref{eq13}).
On a fixed line $r+s=c$ the ee value varies from 
$\phi_1=-1$ at a point $(r,s)=(0, c)$ to $\phi_1=1$
 at (c,0),
and the final values of 
$\phi_1$  can be anything between $+1$ and $-1$.
In order to obtain the final value of ee,
one has to solve the evolution (\ref{eq09}) and especially
to determine the trajectory of the dynamical flow in $r-s$ phase space.

\subsection{Flow Trajectory in a $r-s$ Phase Space}

As the first step to analyze flow trajectories in a phase space,
we consider the simplest case when
reactions such as 
spontaneous production 
(\ref{eq01}), linearly autocatalytic (\ref{eq02})
or quadratically autocatalytic (\ref{eq04}) reactions are active, 
 respectively.
Then the rate equations are simplified as
\begin{align}
\frac{d r}{dt} = f(r)a,
\nonumber  \\
\frac{d s}{dt} = f(s)a ,
\label{eq15}
\end{align}
with the rate coefficient $f$ defined in Eq.(\ref{eq10}).
Flow trajectories are obtained by integrating 
a trajectory equation
\begin{align}
\frac{d s}{dr} = \frac{f(s)}{f(r)} .
\label{eq16}
\end{align}
In the following, an initial condition is set as 
$r=r_0, ~s=s_0, ~a_0=c-r_0-s_0$ (or
$q=q_{1,0}=r_0+s_0, ~\phi_1=\phi_{1,0}=(r_0-s_0)/q_{1,0},~ a_0=c-q_{1,0}$), 
and we discuss how the final ee
value $\phi_{1,\infty}$ depends on the initial value $\phi_{1,0}$ 
for a few typical cases.

\begin{figure}[h]
\begin{center} 
\includegraphics[width=1.\linewidth]{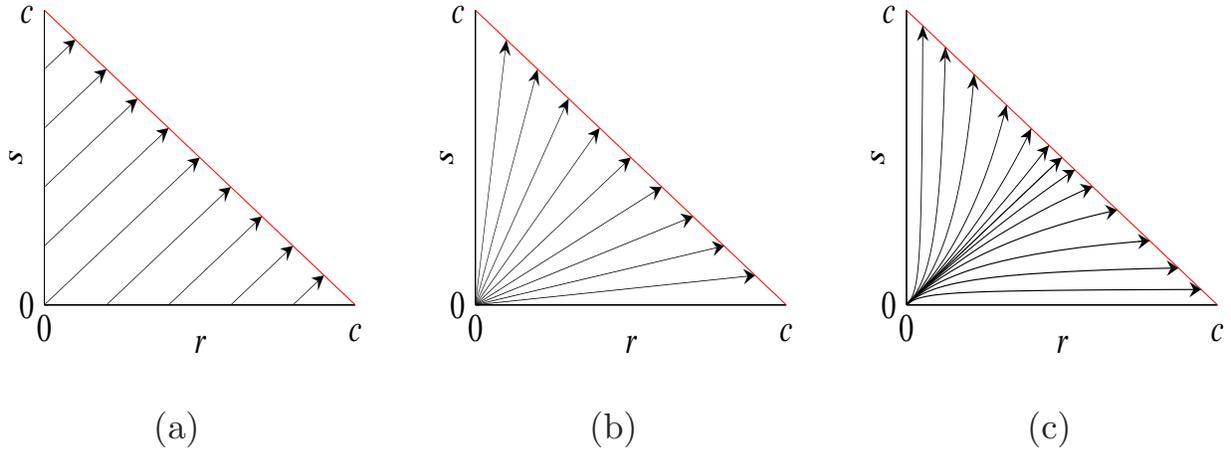}
\end{center} 
\caption{Flow diagrams for (a) 
spontaneous production
($k_0 >0,~k_1=k_2=0$), 
(b) linearly autocatalytic ($k_1>0,~k_0=k_2=0$),
and (c) quadratically autocatalytic ($k_2>0,~k_0=k_1=0$) reactions.
}
\label{fig1}
\end{figure}

\subsubsection{ Spontaneous Production}

\begin{figure}[h]
\begin{center} 
\includegraphics[width=1.\linewidth]{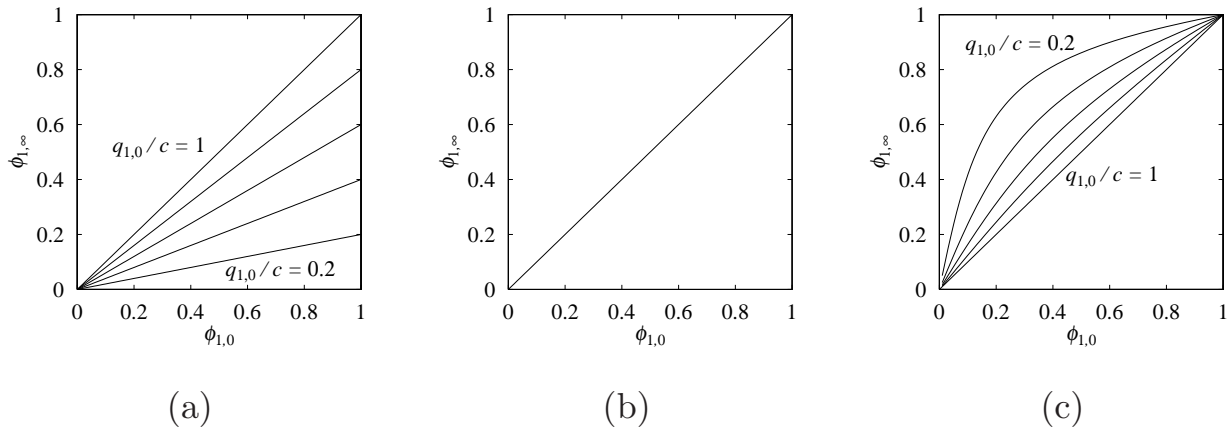}
\end{center} 
\caption{The final ee $\phi_{1, \infty}$ versus the initial ee $\phi_{1,0}$ 
for (a) 
spontaneous production
($k_0 >0,~k_1=k_2=0$), 
(b) linearly autocatalytic ($k_1>0,~k_0=k_2=0$),
and (c) quadratically autocatalytic ($k_2>0,~k_0=k_1=0$) reactions.
Each curve corresponds to a fixed 
ratio of the initial amount of chiral species  
to the total amount of active reactants  $q_{1,0}/c$. As the ratio
decreases, the initial slope
of the curve
decreases in (a), does not change in (b), and increases in (c). 
}
\label{fig2}
\end{figure}

When chiral molecules are produced only spontaneously
and the system has no
autocatalytic processes ($k_0 > 0$, $k_1=k_2=0$),
the trajectory in the $r-s$ phase space is obtained as
\begin{align}
r-s=r_0-s_0=\mbox{const.}
\label{eq17}
\end{align}
The flow diagram is a straight line with a unit slope,
terminating at the fixed line $r+s=c$,
as shown in Fig.\ref{fig1}(a).
Half of the substrate $a_0$ initially present convert to $R$
 and the remaining half to $S$, and 
the ee decreases to the value $\phi_{1,\infty}=(q_{1,0}/c) \phi_{1,0}$. 
The smaller the ratio $q_{1,0}/c$ 
of the initial amount of chiral species  
to the total amount of active reactants  is,
the  more the ee decreases, as shown in Fig.\ref{fig2}(a).
\\

\subsubsection{ Linear Autocatalysis}

When only the linearly autocatalytic production of chiral molecules is active
 ($k_1>0,~k_0=k_2=0$), 
the trajectory is obtained as
\begin{align}
\frac{ s}{r} = \frac{s_0}{r_0}=\mbox{const.}
\label{eq18}
\end{align}
and the flow diagram is a straight line radiating from the origin and
terminating at the fixed line $r+s=c$, 
as shown in Fig.\ref{fig1}(b).
The ee value remains constant as
\begin{align}
\phi_{1,\infty}=\phi_{1,0}=\mbox{const.}
\label{eq19}
\end{align}
and is independent of the initial ratio  $q_{1,0}/c$, 
as shown in Fig.\ref{fig2}(b).

If the erroneous linear catalysis (\ref{eq03}) is at work
in addition to the linear autocatalysis (\ref{eq02}),
 rate equations are modified as
\begin{align}
\frac{d r}{dt} = (k_1r+k_1's)a,
\nonumber  \\
\frac{d s}{dt} = (k_1s+k_1'r)a .
\label{eq20}
\end{align}
These equations can be combined to yield 
\begin{align}
\frac{d \phi_1}{dq_1} = -(1-f)
\frac{\phi_1}{q_1} ,
\label{eq21}
\end{align}
with a fidelity factor $f$ defined by
\begin{align}
f = \frac{k_1-k_1'}{k_1+k_1'}
\label{eq22}
\end{align}
which is unity without the error $k'_1=0$ 
and vanishes when $k_1=k_1'$.
The ee value $\phi_1$ is easily integrated as
\begin{align}
\phi_{1,\infty}=\phi_{1,0} \Big( \frac{q_{1,0}}{c} \Big)^{1-f}
\le \phi_{1,0}.
\label{eq23}
\end{align}
This result shows that the final value 
is less than the initial $\phi_{1,0}$ whenever there is a finite 
error process $f<1$. 
In particular, if $f=0$, the final ee behaves the same with that
of the spontaneous production.

\subsubsection{ Quadratic Autocatalysis}

\begin{figure}
\begin{center} 
\includegraphics[width=0.3\linewidth]{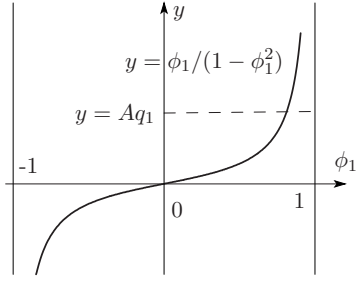}
\end{center} 
\caption{The ee value $\phi_1$ is given at a cross point with two curves,
$y=\phi_1/(1-\phi_1^2)$ and $y=Aq_1$.}
\label{fig3}
\end{figure}

With only a quadratic autocatalysis ($k_2 >0, ~k_0=k_1=0$),
the trajectory is obtained as
\begin{align}
\frac{ 1}{r}-\frac{ 1}{s} = 
\frac{ 1}{r_0}-\frac{ 1}{s_0} = 
\mbox{const.}
\label{eq24}
\end{align}
and the flow diagram is hyperbolae radiating out from the origin and
terminating at the fixed line 
$r+s=c$,
as shown in Fig.\ref{fig1}(c).
Below the diagonal $r=s$ line, the flow bents  
downward and  the ratio $s/r$ decreases as time passes,
and consequently the ee value $\phi_1$ increases.
In other words, 
by using the relation $r=q_1(1+\phi_1)/2$ and
$s=q_1(1-\phi_1)/2$, the ee value is easily found from Eq.(\ref{eq24}) 
to satisfy a relation 
\begin{align}
\frac{\phi_1}{1-\phi_1^2} = 
{A}{q_1}
\label{eq25}
\end{align}
with a constant $A$ determined by an initial condition.
For a given value of $q_1$, the ee value $\phi_1$
 is determined graphically
as a cross point of two curves representing both sides of Eq.(\ref{eq25}), 
as shown in Fig.\ref{fig3}.
Then, it is evident that $\phi_1$ increases as $q_1$ increases.
The asymptotic value corresponding to $q_1=c$ is 
obtained as
\begin{align}
\phi_{1,\infty}=\frac{1}{2cA} \Big[
\sqrt{1+(2 cA)^2} -1 \Big]
\label{eq26}
\end{align}
The relation between the initial value $\phi_{1,0}$ of the ee and 
its final value $\phi_{1,\infty}$ is depicted in Fig.\ref{fig2}(c)
for fixed values of the initial amount $q_{1,0}$ of chiral species
relative to the total amount $c$.
The smaller the initial ratio $q_{1,0}/c$, the more
prominent the ee amplification.

The ee amplification curves possess a remarkable property such that 
they depend on the initial ratio $q_{1,0}/c$ but
are independent of the rate coefficient $k_2$. 
The latter determines the
 time evolution of $\phi_1$ but not its final value.
If there are other processes involved, for instance the spontaneous 
production $k_0 \ne 0$,
then amplification curves should depend on the ratio $k_0/k_2c^2$.

If there is a racemizing 
cross catalysis (\ref{eq05}) with $k_2'$,
the trajectory equation is rewritten as
\begin{align}
\frac{d \phi_1}{dq_1} =\frac{\phi_1( 1-\phi_1^2)}{f_{\times}^{-1}+\phi_1^2}
\frac{1}{q_1}
\label{eq27}
\end{align}
with a crossing effect factor defined by
\begin{align}
f_{\times}= \frac{k_2-k_2'}{k_2+k_2'} 
\label{eq28}
\end{align}
which is unity without 
a cross catalysis term ($k_2'=0$) and vanishes when $k_2=k_2'$.
The solution of Eq.(\ref{eq27}) is given by
\begin{align}
 \frac{\phi_1}{(1-\phi_1^2)^{(1+f_{\times})/2}} = 
A q_1^{f_{\times}} .
\label{eq29}
\end{align}
From this result one can see that the
asymptotic value $\phi_{1,\infty}$ decreases as $f_{\times}$ decreases.
In particular, when $f_{\times}=0$, $\phi$ is independent of $q_1$
and does not vary; $\phi_{1,\infty}=\phi_{1,0}$. Namely,
chirality amplification by $k_2$ is completely cancelled out by the
racemization effect caused by  $k_2'$.

\section{Amplification with Homodimer Catalyst}

In order to understand molecular mechanisms
how the quadratic autocatalysis is brought about,
the concept of dimer catalyst introduced by Kagan and coworkers
may be relevant \cite{girard+98,sato+01,blackmond+01,buhse03}.
Assume that monomers $R$ and $S$ react to form homodimers $R_2$ and $S_2$ 
with a rate $\nu_{\mathrm{hom}}$ and a heterodimer $RS$ with a rate $\nu_{\mathrm{het}}$ 
\begin{align}
R+R \stackrel{\nu_{\mathrm{hom}}}{\rightarrow} R_2, \qquad
S+S \stackrel{\nu_{\mathrm{hom}}}{\rightarrow} S_2 , \qquad
R+S \stackrel{\nu_{\mathrm{het}}}{\rightarrow} RS .
\label{eq30}
\end{align} 
Corresponding decomposition processes are described as
\begin{align}
R_2 \stackrel{\nu_{\mathrm{hom}}'}{\rightarrow} R+R, \qquad
S_2 \stackrel{\nu_{\mathrm{hom}}'}{\rightarrow} S+S , \qquad
RS \stackrel{\nu_{\mathrm{het}}'}{\rightarrow} R+S . 
\label{eq31}
\end{align}
Homodimers thus formed are assumed to 
catalyze of their  own enantiomeric
type as
\begin{align}
A+R_2 \stackrel{k_{\mathrm{hom}}}{\rightarrow} R+R_2, \qquad
A+S_2 \stackrel{k_{\mathrm{hom}}}{\rightarrow} S+S_2, 
\label{eq32}
\end{align} 
whereas heterodimers have no preference in enantioselectivity as
\begin{align}
A+RS \stackrel{k_{\mathrm{het}}}{\rightarrow} R+RS, \qquad
A+RS \stackrel{k_{\mathrm{het}}}{\rightarrow} S+RS
\label{eq33}
\end{align} 

Now, the state of the system is described not only by the concentrations
 of monomers $r$ and $s$, but also by those of dimers 
denoted as  $[R_2]$, $[S_2] $ and $[RS]$ respectively.
The rate equations are
\begin{align}
\frac{d r}{dt} &= k_0a + k_{\mathrm{hom}}a[R_2]
-2 \nu_{\mathrm{hom}} r^2+2 \nu_{\mathrm{hom}}' [R_2]
\nonumber \\
& \quad -\nu_{\mathrm{het}} rs + \nu_{\mathrm{het}}' [RS] +k_{\mathrm{het}}a[RS],
\nonumber  \\
\frac{d [R_2]}{dt} &= \nu_{\mathrm{hom}} r^2- \nu_{\mathrm{hom}}' [R_2],
\nonumber  \\
\frac{d [RS]}{dt} &= \nu_{\mathrm{het}} rs- \nu_{\mathrm{het}}' [RS],
\label{eq34}
\end{align}
and the corresponding ones for the $S$ enantiomer.
The ee or the chiral order parameter is now defined by
\begin{align}
\phi=
\frac{r+2[R_2]-s-2[S_2]}{q} .
\label{eq35}
\end{align}
where $q$ is the total concentration of chiral molecules given as
\begin{align}
q={r+s+2[R_2]+2[RS]+2[S_2]} .
\label{eq36}
\end{align}

If the dimerization and decomposition proceed very fast,
the dimer concentrations take quasi-equilibrium values 
which satisfy $[\dot R_2]=[\dot {RS} ]=[\dot S_2 ]=0$ and
are
determined by the 
instantaneous 
 monomer concentrations $r$ and $s$ as
\begin{align}
[R_2]=\frac{\nu_{\mathrm{hom}}}{\nu_{\mathrm{hom}}'} r^2, \quad
[RS]=\frac{\nu_{\mathrm{het}}}{\nu_{\mathrm{het}}'} rs, \quad
[S_2]=\frac{\nu_{\mathrm{hom}}}{\nu_{\mathrm{hom}}'} s^2 .
\label{eq37}
\end{align}
Then, the equations for monomers in Eq.(\ref{eq34}) 
reduce to Eq.(\ref{eq09})
with the coefficients $k_2$ and $k_2'$ given as
\begin{align}
k_2=k_{\mathrm{hom}} \frac{\nu_{\mathrm{hom}}}{\nu_{\mathrm{hom}}'} , \quad
k_2'=k_{\mathrm{het}} \frac{\nu_{\mathrm{het}}}{\nu_{\mathrm{het}}'} 
\label{eq38}
\end{align}
and with $k_1=k_1'=0$.

\begin{figure}[h]
\begin{center} 
\includegraphics[width=0.9\linewidth]{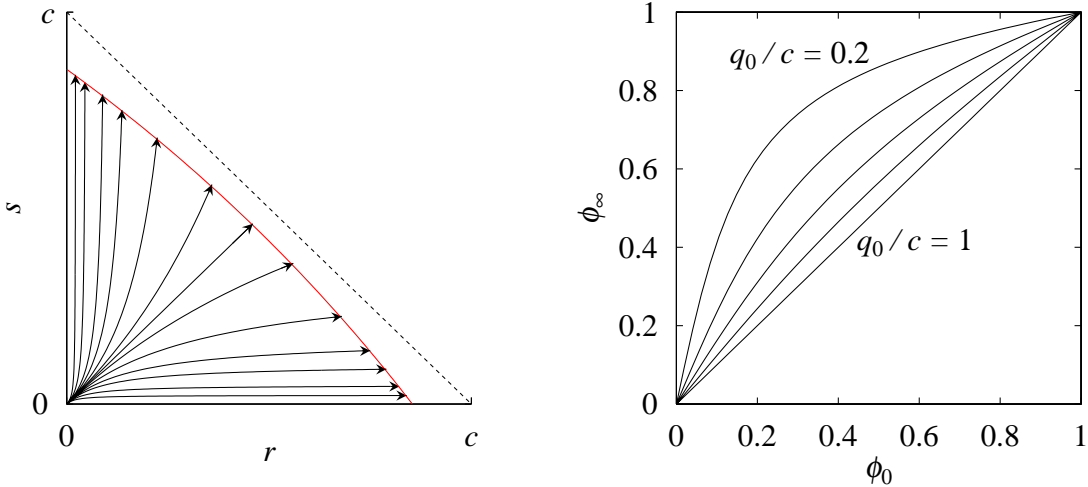}
\end{center} 
\caption{(a) Flow diagrams for homodimer catalyst with $\nu_{\mathrm{hom}} c/\nu_{\mathrm{hom}}'=0.1$ and 
$\nu_{\mathrm{het}}=\nu_{\mathrm{het}}'=0$.
Flow terminates on a curve of fixed points $q=c$.
(b) The final ee value $\phi_{\infty}$ as a function of its initial value 
$\phi_0$ for various amount of initial chiral species $q_0$.
The less $q_0$ is, the more amplified $\phi_{\infty}$ is.
}
\label{fig4}
\end{figure}

As the reaction proceeds, the whole achiral substrate is ultimately
transformed to chiral products  so that 
$a=c-q_{\infty}=0$. 
This represents a quadratic curve of fixed points in $r-s$ phase space.
When there is solely a homodimer catalytic effect $k_2 >0$ without
spontaneous production nor heterodimer racemization $k_0=k_2'=0$, 
the initial state with $(r_0,s_0)$ follows the
hyperbolic flow trajectory (\ref{eq24}) and ends up on the 
curve of fixed points, $q=c$,
as shown in Fig.\ref{fig4}(a).

The ratio $r/s$ increases along the flow trajectory
if it is initially larger than unity. Thus the $r-s$ asymmetry increases,
and the final value of the ee 
$|\phi_{\infty}|$
is larger than the initial value
$|\phi_0|$.
Figure \ref{fig4}(b)
shows $\phi_0-\phi_{\infty}$ relation for cases of positive $\phi$.
The enhancement is more evident when the initial ratio
 $q_0/c$ is smaller.
These essential features are the same with the monomeric system.
 
\section{Amplification with Antagonistic Heterodimer }

Another model for the chiral asymmetry amplification
is a variant of the Frank model
in a closed system
\cite{islas+05}.
In its simplest form, the model consists of reaction schemes of a
spontaneous production (\ref{eq01}), a linear autocatalysis 
(\ref{eq02}) and a heterodimer formation (\ref{eq31}).
Then the rate equations for monomers are the same 
with those of the original Frank model
\cite{frank53}, but they are supplemented with the 
heterodimer formation
\begin{align}
\frac{d r}{dt} &= k_0a + k_1ar -\nu_{\mathrm{het}} rs , 
\nonumber \\
\frac{d s}{dt} &= k_0a + k_1as -\nu_{\mathrm{het}} rs 
\nonumber  \\
\frac{d [RS]}{dt} &= \nu_{\mathrm{het}} rs
\label{eq39}
\end{align}

\begin{figure}[h]
\begin{center} 
\includegraphics[width=1.\linewidth]{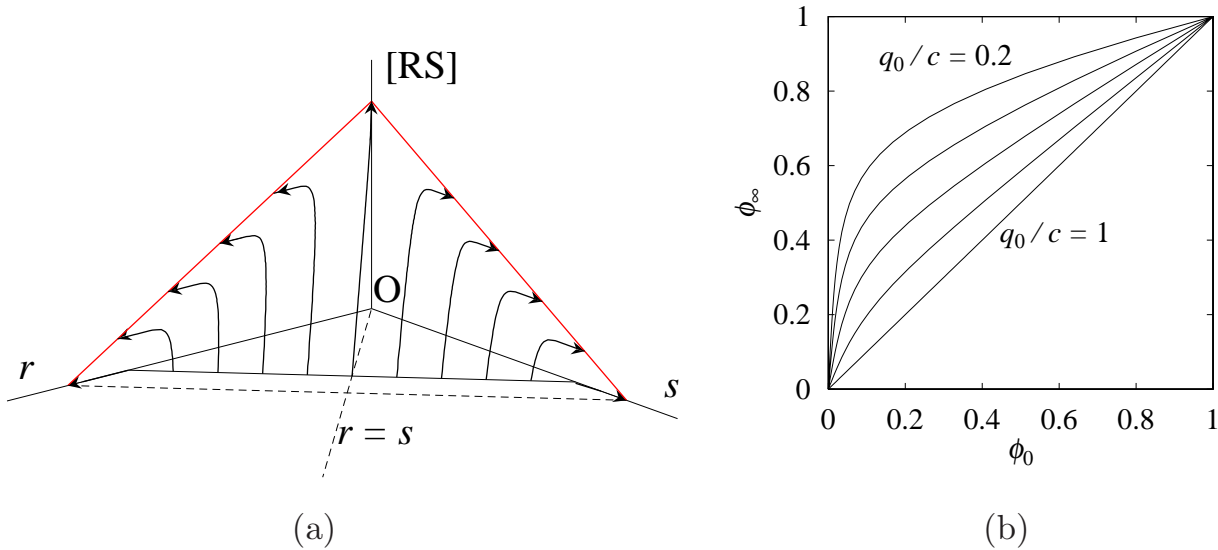}
\end{center} 
\caption{(a) 3D flow diagrams for reactions with antagonistic heterodimers 
$RS$. 
Red lines represent fixed lines.
Initial amount of achiral substrate is $0.2c$, and the remaining 
$q_0=0.8c$ is shared by $R$ and $S$ monomers. There are no heterodimers
initially. Rate coefficients 
 are $k_0=0$, $k_1c=1$, $\nu_{\mathrm{het}}c=10$.
(b) Final ee $\phi_{\infty}$ versus its initial value $\phi_0$ for
a given initial ratio $q_0/c$.
 As $q_0/c$ decreases, the ee amplification increases.
}
\label{fig5}
\end{figure}

Since every achiral substrate is consumed up eventually
$a(t=\infty)=0$ and all the reactions stop asymptotically, 
Eq.(\ref{eq39}) tells us that
the product $rs$ should vanish. 
If there is more $R$ than $S$ initially, 
$S$ monomer disappears ultimately, for instance. 
But $S$ molecules do not disappear nor 
decomposed back into achiral substrate.
They are only incorporated into the heterodimer $RS$.
The system is not determined solely by monomer concentrations 
$r$ and $s$, 
(or $\phi_1$ and $q_1$)
but it depends on heterodimer concentration $[RS]$ as well.
The flow takes place in a three-dimensional phase space of
$r,~s,~[RS]$, as shown in Fig.\ref{fig5}(a).

The asymptotics discussed above forces both the concentrations of
the substrate $a=c-r-s-2[RS]$ and the product $rs$
vanish ultimately.
These two conditions
define a line of fixed points in the three-dimensional $r-s-[RS]$ phase
space.
If the initial state has a prejudice to $R$ enantiomer such as
$r_0 > s_0$, then
the system ends up on a fixed line $r+2[RS]=c$ on a $s=0$ plane,
as shown in Fig.\ref{fig5}(a).
Otherwise with $r_0 < s_0$, the system flows to another fixed line
$s+2[RS]=c$ on a $r=0$ plane.
The plane $r=s$ is a dividing surface that determines the
final preference of chirality.

One can readily find that ee is amplified by the reaction involving
antagonistic heterodimer, as shown in Fig.\ref{fig5}(b).
For a given amount of initial total concentration of chiral species
$q_0=r_0+s_0$, the ee value $\phi$ increases by the completion of a reaction.
No heterodimer is assumed to exist initially, $[RS]_0=0$.
The amplification is more significant for 
smaller amount of $q_0/c$ and $\phi_0$.
We note that the rate constant of heterodimerization $\nu_{\mathrm{het}}$ 
has to be large in order to
produce strong ee amplification.
Minority enantiomer should be quickly incorporated into heterodimers
before the achiral substrate is consumed up.
Otherwise, both enantiomers consume up the achiral resource $A$
and slowly forms heterodimers afterwards
 but without significant ee amplification.
For a large $\nu_{\mathrm{het}}$, minority enentiomers are 
incorporated into heterodimers in an early stage.
Therefore, flow trajectories first point upwards in Fig.\ref{fig5}(a),
indicating the rapid increase of heterodimer concentration $[RS]$.
The trajectory is in fact 
inclined 
 such that the minority enantiomer
decreases its monomeric concentration. When it almost disappears, then the 
trajectory makes a quick turn in the vicinity of 
$r=0$ plane or $s=0$ plane,
 indicating the concentration growth of the surviving majority enantiomer. 

\section{Recycling and Flow: Chiral Symmetry Breaking}

\begin{figure}[h]
\includegraphics[width=0.98\linewidth]{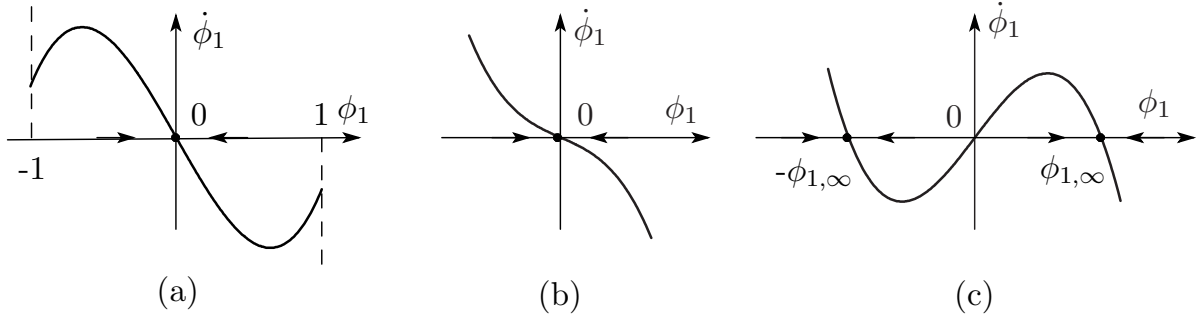}
\caption{$\phi_1$ versus the velocity $\dot \phi_1=A \phi_1-B \phi_1^3$ 
for three typical cases:
(a) $A<B<0$, (b) $A<0<B$ , and (c) $0<A<B$.}
\label{fig6}
\end{figure}

So far, we have discussed several models for the ee amplification. 
There, the irreversibility of the dynamical system is essential
to make the final state dependent on the initial condition.
In this section,
we consider the effect induced by introducing reversibility in the system
by including various types of recycling processes.
For simplicity, we consider in terms of a monomeric system.

In addition to the catalytic production processes (\ref{eq01}) to
(\ref{eq05}), we include the recycling process
such that enantiomeric products $R$ or $S$ 
undergo a back reaction 
to the substrate $A$ 
\cite{saito+04a}
with a reaction rate $\lambda$ as
\begin{align}
R \stackrel{\lambda}{\rightarrow} A, \qquad
S \stackrel{\lambda}{\rightarrow} A .
\label{eq40}
\end{align} 
We call this reaction a linear recycling hereafter.
One can also assume a nonlinear type of recycling process
\cite{saito+05c} 
 such that different chiral species by their encounter react
 back to achiral molecules as
\begin{align}
R+S \stackrel{\mu}{\rightarrow} 2A.
\label{eq41}
\end{align} 
This back reaction can be a combined effect of 
the following two processes 
\begin{align}
R+S \stackrel{\mu}{\rightarrow} R+A \nonumber\\
R+S \stackrel{\mu}{\rightarrow} S+A.
\label{eq41a}
\end{align} 
The rate equations now read as
\begin{align}
\frac{d r}{dt} = [f(r)+k_1's+k_2'rs]a - \mu rs - \lambda r
\nonumber  \\
\frac{d s}{dt} = [f(s)+k_1'r+k_2'rs]a - \mu rs - \lambda s .
\label{eq42}
\end{align}
The term with $\mu$ 
has a same form with the contribution of Frank's
mutual inhibition \cite{frank53}.
The corresponding time-evolution of $\phi_1$ is readily derived in the
form of Eq.(\ref{eq13}) 
with coefficients $A$ and $B$ determined by symmetric quantities
$q_1$ and $a=c-q_1$ as
\begin{align}
A(t)& =-2\frac{a}{q_1}(k_0+ k_1' q_1)+B(t), 
\nonumber \\
B(t)&=\frac{1}{2}[(k_2-k_2')a+\mu ]q_1.
\label{eq43}
\end{align}
The evolution Eq.(\ref{eq13}) of the order parameter has a similar form
with the time-dependent Landau equation
\cite{landau+54}
 which is fundamental in nonequilibrium phase transitions.
The asymptotic value of the order parameter $\phi_{1, \infty}$ is
determined as the zero of the velocity $\dot \phi_1$.
 The main difference from the standard model of phase transitions lies in
the time-dependence in the coefficients $A(t)$ and $B(t)$ induced by
that of the achiral concentration $a(t)$ and the total chiral concentration
$q_1(t)$.
Because the concentrations $a$ and $q_1$ are non-negative, 
$A(t)$ cannot exceed $B(t)$; $A(t)\le B(t)$.

In previous sections we have 
focused our studies on the cases with $\lambda=\mu=0$.
In these cases, the asymptotic value of $a$ vanishes, 
$a(t=\infty)=0$ so that $A(t=\infty)=B(t=\infty)=0$, and
the asymptotic value of the order parameter $\phi_1$ cannot be determined from
Eq.(\ref{eq13}).
Another example with $A=B=0$ happens for a special case with
$k_0=k_1'=\mu=0$ and $k_2=k'_2$. In this case, 
we can calculate trajectories and find a fixed line shifted from $a=0$
to the one with a finite value of $a(t=\infty)$ \cite{saito+05c}: 
Flow trajectories are along lines passing through the origin, 
similar to those shown in Fig.\ref{fig1}(b), 
but terminating at a shifted fixed line. 
Since $a=0$ is no more a fixed line, trajectories  not only point
 out from the origin, but also come out from the line $a=0$.
 
In all the other cases,
with either a linear or a nonlinear recycling process or with both, 
the coefficients $A$ and $B$ are no longer zero at the same time, and
a definite value of the order parameter $\phi_1$ is obtained asymptotically.
The reason for the above is as follows.
If the nonlinear recycling exists  as $\mu >0$, $B$ becomes nonzero since 
$\mu q_{1,\infty} >0$.
If the linear recycling exists as $\lambda>0$, not all the achiral substrate transform to chiral products
 but a finite amount remains asymptotically as $a(t=\infty)>0$. 
 Therefore, nonzero values of $k_0$, $k_1'$ or $k_2 \ne k_2'$ give 
contributions to the coefficients $A$ or $B$.

For nonzero $A$ and/or $B$, the velocity 
$\dot \phi_1=d \phi_1/dt$ has three typical behaviors 
as a function of the order parameter $\phi_1$,
as shown in Fig.\ref{fig6}.
In the region with a positive (negative) velocity $\dot \phi_1$, 
$\phi_1$ increases (decreases) and moves to the right (left), 
as indicated by arrows on the $\phi_1$ axis in Fig.\ref{fig6}.
When $A$ is negative, as in Fig.\ref{fig6}(a) or (b),
$\phi_1$ eventually approaches the racemic value $\phi_1=0$ 
within its range of definition $|\phi_1| \le 1$:
The racemic state is a stable fixed point for $A<0$.
Equation (\ref{eq43}) tells us that 
this happens when the spontaneous production $k_0$ or
error in a linear autocatalytic process $k_1'$ are sufficiently strong,
or quadratic catalysis $k_2$ is weaker than the cross catalysis $k_2'$.
The coefficient of the linear autocatalysis $k_1$ 
as well as the linear recycling process $\lambda$
are absent in the coefficients
$A$ and $B$, and cannot affect directly the chirality of the system.
These linear processes with $k_1$ and $\lambda$
affect the chirality only implicitly through $a$ and $q_1$.

When $A$ is positive, as in the case of Fig.\ref{fig6}(c), 
the coefficient  of the cubic term $B$ is also positive, 
and the velocity $\dot \phi$ vanishes at three values of $\phi_1$
in the range of $-1\le \phi_1 \le 1$.
This is possible 
if a strong quadratic autocatalysis $k_2 > k_2'$ exists 
together with a linear recycling $\lambda>0$,
or if a linear autocatalysis $k_1$ and a nonlinear recycling $\mu>0$ coexist.
By following 
the direction indicated by the arrows
 for positive $\phi_1$,
the order parameter ends up at a definite value
\begin{align}
\phi_{1,\infty}= \sqrt{\frac{A(t=\infty)}{B(t=\infty)}}
\label{eq44}
\end{align}
and the chiral symmetry breaking takes place.
If the system starts with a negative $\phi_1$, it stops at
$- \phi_{1,\infty}$. 
The final state of the system depends on the initial sign of 
$\phi_{1,0}$, 
but independent of its magnitude $|\phi_{1,0}|$.
We note that the asymptotic values of the coefficients $A(t=\infty)$ and $B(t=\infty)$ are
independent of the initial condition and so is $|\phi_{1,\infty}|$.
This situation is what we mean the "chiral symmetry breaking" in the sense of
statistical mechanics.

\begin{figure}[h]
\begin{center} 
\includegraphics[width=1.\linewidth]{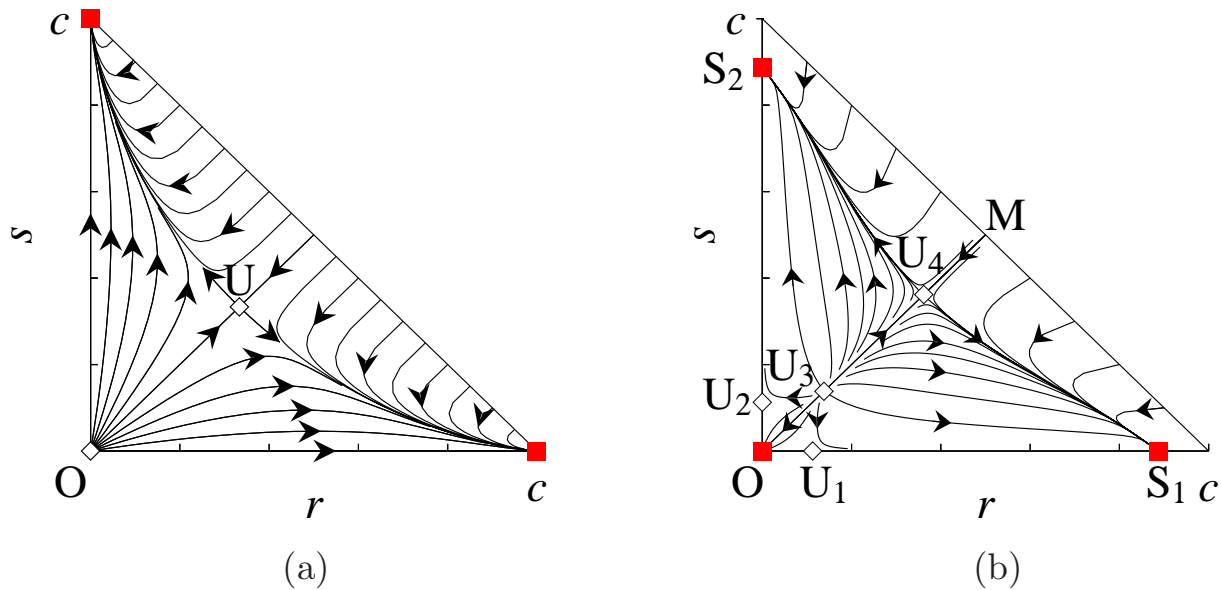}
\end{center} 
\caption{Flow diagrams with (a) a
 linear autocatalysis and a nonlinear recycling
( $k_1=\mu>0,~k_0=k_2=\lambda=0$ ) 
 similar to the Frank model,  and (b) a 
quadratic autocatalysis and a linear 
recycling ($k_2c^2=\lambda>0,~k_0=k_1=\mu=0$).
}
\label{fig7}
\end{figure}

We first discuss the chiral symmetry breaking
when there is a nonlinear recycling, $\mu>0$
\cite{frank53,saito+05c}.
The rate Eq.(\ref{eq42}) tells us that $a=c-r-s=0$ is no more a fixed line,
but we have many fixed points.
The simplest is the case with only a linear autocatalysis
$k_1>0$ but nothing else ($k_0=k_1'=k_2=k_2'=\lambda=0$).
In this case we have four fixed points in the $r-s$ phase space: 
two unstable ones, O and U,
and two stable ones at $(r,s)=(c,0)$ or $(0,c)$,
as is shown in Fig.\ref{fig7}(a).
At the stable fixed points the final ee take values 
$\phi_{1,\infty}=\pm 1$.
These values agree with those expected from Eq.(\ref{eq44}) with 
$A(t)=B(t)=\mu q_1/2$ as determined from Eq.(\ref{eq43}).
The nonlinear recycling (42) returns the same amount of enantiomers back to the achiral substrate, but the relative decrement in concentration is larger for the minority enantiomer than for the majority one. The damage is further amplified by a large reduction of linear autocatalytic effect for the minority enantiomer, and it disappears at last.

As for the second example, we consider the case 
with a quadratic autocatalysis $k_2>0$ and a linear recycling $\lambda>0$.
Because the linear recycling takes place whenever there is nonzero concentrations of enantiomers, a few achiral substrate always remain, $a(t=\infty)\ne 0$.
Thus the diagonal line $q_1=r+s=c$ is no more a fixed line.
Instead there appear fixed points in general.
In the present case with a finite $\lambda>0$,
totally seven fixed points appear; three stable (O, S$_{1,2}$) 
and four unstable ones (U$_{1-4}$), as shown in Fig.\ref{fig7}. 
As reaction proceeds, the 
system approaches a state associated with one of 
the stable fixed points. The origin O is not interesting,
since all the chiral products are recycled back to the achiral substrate.
Also its influence extends only in a small region around O, 
bounded by U$_{1-3}$, 
if the life time $1/\lambda$ of chiral products is long enough.
Close to this fixed point, 
influence of the spontaneous production process with $k_0$
should be taken into account, anyhow. 

When the initial system is sufficiently away from O, the system approaches
S$_1$ or S$_2$; Both correspond to homochiral states with 
$\phi_{1,\infty}=\pm 1$.
When the initial configuration is close to the racemic state
or the diagonal line $r=s$, the system approaches a racemic fixed point 
U$_4$ at first.
But, while the recycling process returns the chiral 
enantiomers back to achiral substrate, the majority enantiomer increases its
population at the cost of the minority one along the flow curve diverging from
U$_4$ to S$_1$
or to S$_2$, and eventually the whole system becomes homochiral.

One may wonder whether recycling processes such as a linear 
or a nonlinear back reaction exist in relevant autocatalytic systems. 
So far, we are not aware of their existence.
It is, however, possible that the back reaction rates $\lambda$ 
or $\mu$ are nonzero but exceedingly small
to be detected in laboratory experiments. 
Concerning with the problem of homochirality in life, 
very small $\lambda$ or $\mu$ are not unimaginable,
 considering the geological time scale for its establishment on earth.
 
On the other hand, a possibility to provide a system with a
recycling process is proposed theoretically \cite{saito+05b}:
One simply let the reaction with  ee amplification run
in an open flow reactor. 
In an open system there is reactant flows between the system
and the environment $E$.
Let the solution with the achiral substrate $A$ be supplied by an inflow
with an influx $F$ as
\begin{align}
E \stackrel{F}{\rightarrow} A
\label{eq45}
\end{align} 
and the products $R$ and $S$ as well as the substrate $A$ 
be taken out by an outflow
 with  a flow out rate $\lambda$ as
\begin{align}
A \stackrel{\lambda}{\rightarrow} E, \qquad
R \stackrel{\lambda}{\rightarrow} E, \qquad
S \stackrel{\lambda}{\rightarrow} E .
\label{eq46}
\end{align} 
In terms of rate equations the processes are described as
\begin{align}
\frac{dr}{dt}&= f(r)a-\lambda r , \quad 
\frac{ds}{dt}= f(s)a-\lambda s , 
\nonumber \\
\frac{da}{dt}&= -[f(r)+f(s)]a+F-\lambda a .
\label{eq47}
\end{align}
Since the total concentration $a+r+s$ follows the time
evolution $d(a+r+s)/dt=F-\lambda (a+r+s)$, 
it approaches the steady state value $F/\lambda$ with a relaxation time 
$1/\lambda$.
This is a consequence of unbiased outflow (\ref{eq47}) of all reactants
with the same rate $\lambda$.
Consequently, even though we are dealing with an open system under a flow,
the analysis is similar to the closed system by replacing the
total concentration $c$ 
with the steady state value $F/\lambda$.
Instead of recycling, therefore, constant supply of the substrate 
allows the system to reach a certain fixed point with a definite
value of the order parameter $\phi_1$, independent of the initial
condition.

\section{Summary and Discussions}

\begin{figure}
\begin{center} 
\includegraphics[width=0.3\linewidth]{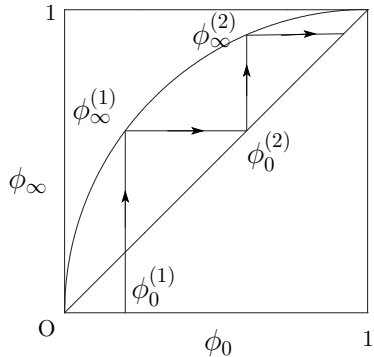}
\end{center} 
\caption{Schematics of enantiomeric excess amplification.}
\label{fig8}
\end{figure}

We have presented various simple scenarios that we are aware of in
relevance 
 to the ee amplification of the Soai reaction;
a quadratic autocatalytic model in a monomer or a homodimer system, 
or a linear autocatalytic model in an antagonistic heterodimer system.
All these models can realize ee amplification such that the final
value of the ee $|\phi_{\infty}|$ depends on but is 
larger than the initial value $|\phi_0|$,  
as schematically shown in Fig.\ref{fig8}.
The curve in the figure
represents $\phi_{0}-\phi_{\infty}$ relation for a given 
initial ratio $q_0/c$, namely the ratio of the 
amount $q_0$ of the total chiral initiators $R$ and $S$ relative to 
that of the total reactants $c$.
Amplification is more enhanced if the ratio of the chiral initiator
$q_0/c$ is smaller.
This plot also shows the possibility to increase the final ee by repeating
the reaction.

To achieve high ee values, Soai et al. performed experiments of
consecutive autocatalytic reactions, where after each run of the reactions,
resulting solutions are quenched by adding acid and purified and
reused in the next reactions. 
How the high ee value is attained in this type of experimental procedures
can be explained illustratively in Fig.8.
By choosing a specified initial condition $\phi_0^{(1)}$, the autocatalytic
reaction ends up giving a final ee as $\phi=\phi_\infty^{(1)}$ determined
by the intersection between the vertical line starting from $\phi_0^{(1)}$
point and the round curve.
The consecutive reaction starts with the purified products with the ee
$\phi_\infty^{(1)}$ as the initial ee $\phi_0^{(2)}$.
This feature is illustrated by extending a horizontal line from 
$\phi^{(1)}_\infty$ on the round curve to the $\phi_0=\phi_\infty$
straight slope.
This second reaction yields the ee value $\phi_{\infty}^{(2)}$, which is determined
 by the intersection between the vertical line starting from 
 $\phi_0^{(2)}=\phi_\infty^{(1)}$ point and the round curve.
 To repeat these procedures, the ee approaches rapidly 
 the perfect $|\phi_\infty^{(\infty)}|=1$ with the speed depending on the curvature of the curve, which is determined by $q_0/c$.
This is a mechanism of ee amplification and consecutive autocatalytic 
reactions, 
in contrast with the chiral symmetry breaking in the sense of statistical 
 mechanics, where the final ee is unique and is reached by 
 a single procedure.

In all the cases considered, autocatalytic processes must be present, 
whether it be linear or nonlinear.
To understand the actual mechanism of autocatalysis 
for the Soai reaction, the identification of
the process at a molecular level is necessary, but that is out of
scope of the present review.

We would like to point out 
 another interesting feature,  namely  the spatial aspect of the reaction
\cite{saito+04b,brandenburg+04,shibata+06}.
In usual experiments, the reactor is stirred 
to keep the system homogeneous.
If there is no stirring, 
inhomogeneous patterns may develop between 
domains of two enantiomers.
Analysis of these patterns 
may give some insight into the mechanism of reaction dynamics.
Let us imagine, for instance, a time evolution of patterns of chiral domains
for diffusionless systems with ee amplification and with a
chiral symmetry breaking.
In an initial stage chiral domains are nucleated randomly, and the domain
pattern may look similar for both systems. 
After domains come in touch with each other, 
they behave differently: 
For a system with ee amplification, domain pattern freezes because 
there remains no achiral substrate for further reactions.
On the other hand, for a system with  the chiral
symmetry breaking, neighboring domains compete with each other at 
domain boundaries to extend
their own territories, and the domain pattern evolves to selection slowly.
When diffusion or hydrodynamic flow effects are taken into account,
the system with ee amplification simply homogenizes 
and the ee value shall be averaged out.
For the system with the chiral symmetry breaking, these transport
processes speed up the chirality selection and enhance the final value of ee
\cite{brandenburg+04,shibata+06}.

So far we have discussed the ee amplification by starting from an initial
state with a finite enantiomeric imbalance, and the effect of spontaneous
reaction is mostly neglected, $k_0=0$.
However, reaction experiments without adding chiral substances
do produce chiral species
\cite{soai+03,gridnev+03,singleton+03},
implying that
 the coefficient $k_0$ of the spontaneous reaction should be
nonzero. 
In these experiments, it is found that 
values of the ee order parameter $\phi$ of many runs are 
 distributed widely between $-1$ and $+1$,
and the probability distribution function has double peaks at
positive and negative values of $\phi$.

There is a theoretical study on the asymptotic shape 
of probability distribution for nonautocatalytic and linearly autocatalytic
systems with a specific initial condition of no chiral enantiomers
\cite{lente04,lente05}.
Even though no ee amplification is expected in these cases,
the probability distribution with a linear autocatalysis has symmetric double
peaks  
at $\phi=\pm 1$
when $k_0$ is far smaller than $k_1$; $k_0 \ll k_1/N$ with
$N$ being the total number of all reactive chemical species, $A$, $R$ and $S$.
This can be explained by the "single-mother" scenario for the realization of
homochirality, as follows.
From a completely achiral state, one of the chiral molecule, say $R$, 
is produced spontaneously and randomly after an average time $1/2k_0N$. 
Then, the second $R$ is produced by the autocatalytic process whereas
for the production of the first $S$  molecule
the spontaneous production is necessary.
The  waiting time for the second $R$ is shorter than that for the first $S$
by a factor about $k_0/k_1$.
If this factor is far smaller than $1/N$, 
there is only a negligible chance for the production of $S$ enantiomer
until all the achiral substrates turn into $R$. 
All the chiral products are the descendants of the first mother $R$
produced by the autocatalysis, and no second mother is born
by the spontaneous production: This is a single-mother scenario.
It strongly reflects the stochastic character of the
chemical reaction, and is not contained in the average description using
 rate equations.
Further studies are necessary as for the stochastic aspects of the
Soai reaction \cite{saito+07}.

In relation to the origin of homochirality in life, 
an interesting question is whether the ee amplification is
sufficient to achieve homochirality.
From the experimental results of the Soai reaction without chiral initiator,
the magnitude of the final ee $|\phi_{\infty}|$ is distributed 
from a small value to a value  close to unity. 
It is rather difficult to imagine
that a value close to unity has been accidentally selected  
on various places on earth in the prebiotic era.
It seems rather natural that chiral symmetry breaking took place
and a unique value of $|\phi|$ has been finally
selected. 
As has been discussed in the previous section,
the flow in the open system can alter the system with
ee amplification to the one with the chiral symmetry breaking.
So we may say that the system with ee amplification in an open flow is 
one possibility.
Recently, there appears several 
theoretical proposals such as polymerization models
\cite{sandars03,brandenburg+05,saito+05b} 
or a polymerization-epimerization model
\cite{plasson+04}
to realize the chirality selection in polymer synthesis.
Although  we are still far away from 
satisfactory understanding of the origin
of homochirality in life, 
the discovery of the autocatalytic reaction by Soai 
gives us great impetus and a sense
of reality to this problem.



\begin{thebibliography}{99}

\bibitem{stryer98}
Stryer L(1998)Biochemistry,  Feeman and Comp, New York

\bibitem{pasteur849}
Pasteur L(1849)Comptes Rendus 28:477

\bibitem{japp898}
Japp FR(1898)Nature 58:452

\bibitem{noyori02}
Noyori R(2002)Angew Chem Int Ed 41:2008

\bibitem{calvin69}
Calvin M(1969)Chemical Evolution.Oxford University Press, 
Oxford

\bibitem{goldanskii+88}
Goldanskii VI, Kuz'min VV(1988)Z Phys Chem (Leipzig) 
269:216


\bibitem{gridnev06}
Gridnev ID(2006)Chem Lett 35:148

\bibitem{peason898}
Pearson K(1898)Nature 58:496 and 59:30
 
\bibitem{mills32}
Mills WH(1932)Chem Ind (London) 51:750
 

\bibitem{soai+95}
Soai K, Shibata T, Morioka H, Choji K(1995)Nature 378:767

\bibitem{soai+00}
Soai K, Shibata T, Sato I(2000)Acc Chem Res 33:382

\bibitem{soai+03}
Soai K, Sato I, Shibata T, Komiya S, Hayashi M, Matsueda Y, Imamura H, 
Hayase T, Morioka H, Tabira H, Yamamoto J, Kowata Y(2003)Tetrahedron:
Asymmetry 14:185

\bibitem{gridnev+03}
Gridnev ID, Serafimov JM, Quiney H, Brown JM(2003)Org Biomol Chem
1:3811

\bibitem{singleton+03}
Singleton DA, Vo LK(2003)
Org Lett 
5:4337

\bibitem{frank53}
Frank FC(1953)Biochimi Biophys Acta 11:459

\bibitem{landau+54}
Landau LD, Khalatnikov IM(1954)Dokl Akad Nauk SSSR 96:469




\bibitem{avetisov+96}
Avetisov V, Goldanskii V(1996)Proc Nat Acad Sci USA 93:11435

\bibitem{girard+98}
Girard C, Kagan HB(1998)Angew Chem Int Ed 37:2922

\bibitem{kondepudi+01}
Kondepudi DK, Asakura K(2001)Acc Chem Res 34:946

\bibitem{todd02}
Todd MH(2002)Chem Soc Rev 31:211

\bibitem{iwamoto02}
Iwamoto K(2002) Phys Chem Chem Phys 4:3975

\bibitem{iwamoto03}
Iwamoto K(2003) Phys Chem Chem Phys 5:3616

\bibitem{saito+04a}
Saito Y, Hyuga H(2004)J Phys Soc Jpn 73:33

\bibitem{saito+04b}
Saito Y, Hyuga H(2004)J Phys Soc Jpn 73:1685

\bibitem{saito+05a}
Saito Y, Hyuga H(2005)J Phys Soc Jpn 74:535

\bibitem{saito+05b}
Saito Y, Hyuga H(2005)J Phys Soc Jpn 74:1629

\bibitem{saito+05c}
Saito Y, Hyuga H(2005)Chirality Selection Models in a Closed System.
In: Linke, AN (ed) Progress in Chemical Physics Research, NOVA, New York, Ch.3 p.65

\bibitem{shibata+06}
Shibata R, Saito Y, Hyuga H(2006)Phy Rev E 74:026117-1

\bibitem{sato+01}
Sato I, Omiya D, Tsukiyama K, Ogi Y, Soai K(2001) 
Tetrahedron: Asymmetry 12:1965

\bibitem{sato+03}
Sato I, Omiya D, Igarashi H, Kato K, Ogi Y, Tsukiyama K, Soai K(2003) 
Tetrahedron: Asymmetry 14:975

\bibitem{blackmond+01}
Blackmond DG, McMillan CR, Ramdeehul S, Shorm A, Brown JM(2001)
J Am Chem Soc 123:10103

\bibitem{buhse03}
Buhse T(2003)Tetrahedron: Asymmetry 14:1055



\bibitem{islas+05}
Islas JR, Lavabre D, Grevy J-M, Lamoneda RH, Cabrera HR,
Micheau J-C, Buhse T(2005)Proc Nat Acad Sci USA 102:13743

\bibitem{lente04}
Lente G(2004)J Phys Chem 108:9475

\bibitem{lente05}
Lente G(2005)J Phys Chem 109:11058


\bibitem{brandenburg+04}
Brandenburg A, Multamaki T(2004)Int J Astrobiol 
3:209

\bibitem{saito+07}
Saito Y, Sugimori T and Hyuga H(2007) to appear.

\bibitem{sandars03}
Sandars PGH(2003)Orig Life Evol Biosph 33:575

\bibitem{brandenburg+05}
Brandenburg A, Andersen AC, H\"ofner S, Nilsson M(2005)Orig Life Evol Biosph 
35:225

\bibitem{plasson+04}
Plasson R, Bersini H, Commeyras A(2004)Proc Nat Acad Sci USA 
101:16733

\end{thebibliography}
\end{document}